\documentclass[twocolumn]{aa}
\DeclareUnicodeCharacter{2212}{-}
\DeclareUnicodeCharacter{0301}{\'{e}}
\usepackage{hyperref}
\usepackage{graphicx}
\usepackage{xcolor}
\usepackage{txfonts}
\begin{document}
    
    \title{A search for mode coupling in magnetic bright points}


   \author{A. Berberyan, 
   \inst{1}
      P. H. Keys, 
   \inst{2}
   D. B. Jess, 
   \inst{2}
\and
   D. J. Christian, 
   \inst{1}    
          }

   \institute{
 Department of Physics and Astronomy, California State University Northridge, Nordhoff St, Northridge, CA 91330, USA \\
  \email{arthur.berberyan.595@my.csun.edu}
   \email{damian.christian@csun.edu}
    \and
Astrophysics Research Centre, School of Mathematics and Physics, Queen’s University, Belfast, BT7 1NN, Northern Ireland, U.K. \\
\email{p.keys@qub.ac.uk},  
\email{d.jess@qub.ac.uk}
             }

   \date{Received June xx,2024; accepted July xx 2024}

 
  \abstract
   {
   Magnetic bright points (MBPs) are one of the smallest manifestations of the magnetic field in the solar atmosphere and are observed to extend from the photosphere up to the chromosphere. As such, they represent an excellent feature to use in searches for types of magnetohydrodynamic (MHD) waves and mode coupling in the solar atmosphere. 
   }
   {In this work, we aim to study wave propagation in the lower solar atmosphere by comparing intensity oscillations in the photosphere with the chromosphere via a search for possible mode coupling, in order to establish the importance of these types of waves in the solar atmosphere, and their contribution to heating the chromosphere.
   }
   {These observations were conducted in July 2011 with the Rapid Oscillations of the Solar Atmosphere (ROSA) and the Hydrogen-Alpha Rapid Dynamics Camera (HARDCam) instruments at the Dunn Solar Telescope. Observations with good seeing were made in the G-band and H$\alpha$ wave bands. 
   Speckle reconstruction and several post facto techniques were applied to return science-ready images. The spatial sampling of the images was 0.069$\arcsec$/pixel (50 km/pixel).
 We used wavelet analysis to identify traveling MHD 
 waves and derive frequencies in the different bandpasses. 
We isolated a large sample of MBPs using an automated tracking algorithm throughout our observations. Two dozen of the brightest MBPs were selected from the sample for further study. 
   }
   {We find oscillations in the G-band MBPs,   
   with frequencies between 1.5 and 3.6 mHz. Corresponding MBPs in the lower solar chromosphere observed in H$\alpha$ show a frequency range of 1.4 to 4.3 mHz. In about 38\% of the MBPs, the ratio of H$\alpha$ to G-band frequencies was near two. Thus, these oscillations show a form of mode coupling where the transverse waves in the photosphere are converted into longitudinal waves in the chromosphere. 
   The phases of the H$\alpha$ and G-band light curves show strong positive and negative correlations only 21\% and 12\% of the time, respectively.}
   {From simple estimates we find an energy flux of $\approx$45 $\times 10^{3}$ W m$^{-2}$ and show that the energy flowing through MBPs is enough to heat the chromosphere, 
   although higher-resolution data are needed to explore this contribution further. 
   Regardless, mode coupling is important in helping us understand the types of MHD waves in the lower solar atmosphere and the overall energy budget. 
   }

               \keywords{
               Sun: oscillations -- 
               Sun: chromosphere --
               Sun: photosphere --  
               Magnetohydrodynamics (MHD) --
                methods: statistical.
            }  

   \titlerunning{A search for mode coupling in magnetic bright points}
   \authorrunning{Berberyan et al.}
   \maketitle
%

\section{Introduction}
How the Sun transports energy through its dynamic lower solar atmosphere is a challenging topic that has been puzzling researchers for decades \citep[e.g., see the recent reviews by][]{2015SSRv..190..103J, 2023LRSP...20....1J}. Significant work in recent years has enabled magnetohydrodynamic (MHD) waves be not only detected in the Sun's lower atmosphere but also studied in detail, notably with regard to their modal composition \citep{2017ApJ...842...59J, 2022ApJ...938..143G, 2023ApJ...954...30A}, embedded energy flux \citep{2022ApJ...930..129B, 2023ApJ...945..154M}, rates of damping over both spatial and temporal domains \citep{2015ApJ...806..132G, 2021RSPTA.37900172G, 2021A&A...648A..77R}, and the characteristics they demonstrate as they traverse solar plasma dominated by magnetic or plasma pressure \citep{2019ApJ...883..179K, 2021A&A...656A..87M, 2024NatCo..15.2667K}. Importantly, propagating MHD waves are believed to carry mechanical energy flux into the solar chromosphere, which they can subsequently dissipate to heat this layer of the atmosphere. 

Of particular interest is what happens to embedded wave modes as they pass through regions of the solar atmosphere where the magnetic and plasma pressures are approximately equal, such as the so-called equipartition region \citep[e.g.,][]{2008SoPh..251..251C, 2018NatPh..14..480G, 2020ApJ...892...49H}. 
MHD wave mode conversion occurs in plasmas where the waves can change their modes as they propagate through mediums. Mode conversion is plausible when the MHD waves pass through regions characterized by density and magnetic field changes.

However, the promising wave heating models (i.e., those that can provide enough energy to heat the solar corona) often overlooked
the potentially important role of heating the chromosphere and thus the need to understand energy transport in this region.  The behavior of small  
magnetic structures called magnetic bright points (MBPs), is important in understanding the energy transport between the photosphere and lower chromosphere.  
MBPs are some of the smallest observable objects in the
photosphere, appearing as intensity enhancements within the
intergranular lanes, and have magnetic field strengths of over one kilogauss \citep{crockett2010tracking}. 
Although capable of carrying energy into the chromosphere, transverse MHD waves cannot easily dissipate their energy without first passing through an intermediary stage.  However, compressible longitudinal modes can readily deposit their energy through shock formation in the solar atmosphere \citep{zhugzhda1995propagation}. Thus, transforming transverse wave modes into longitudinal modes can provide this heating mechanism \citep{kalkofen1997oscillations}. Transverse waves with frequency $\nu_k$ are converted into longitudinal waves at twice this frequency, $\nu_\lambda$ (2$\nu_k$). 

Evidence of this mode-coupling in network bright points (NBPs) has been presented \citep{bloomfield2004wavelet, mcateer2003observational}.  These longitudinal waves are predicted to shock and heat the chromosphere \citep{zhugzhda1995propagation}.  
Studies from \cite{mcateer2003observational}
for mode coupling in the chromosphere were conducted with NBPs taken in the $Ca\, II\, K_3$, Mg$_1b_1$, Mg$_1b_2$, and H$\alpha$ bandpasses.    
Four NBPs were investigated 
and found to show possible transverse longitudinal wave propagation from the photosphere into the lower chromosphere.
It was concluded that these bright points could produce energy transport in the chromosphere through mode coupling.
We note that the McAteer study focused on NBPs found at chromospheric heights as brightenings (especially in Ca II $K_3$) that are spatially coherent with small-scale magnetic elements, whereas here we focus on MBPs in the photosphere, that is, small-scale magnetic elements 
often found within intergranular lanes.

In this work, we searched for observational evidence of
mode coupling by studying oscillations of MBPs observed in the photosphere (G-band) to the chromosphere (H$\alpha$). In Section 2, we present a summary of the Rapid Oscillations of the Solar Atmosphere (ROSA), the Hydrogen-Alpha Dynamics Camera (HARDCam) observations, and our data analysis.
The results are then presented in Sect. 3 and discussed in comparison to previous studies in Sect. 4. We also investigate whether the phases of the observed oscillations are consistent with mode coupling and produce enough energy to heat the chromosphere. Finally, Sect. 5 summarizes our findings and future extensions of this work.
 
 \section{Observations and analysis}
\begin{figure*}[t]
   \centering
   \vspace{-1.4in}
    \includegraphics[width=9cm]{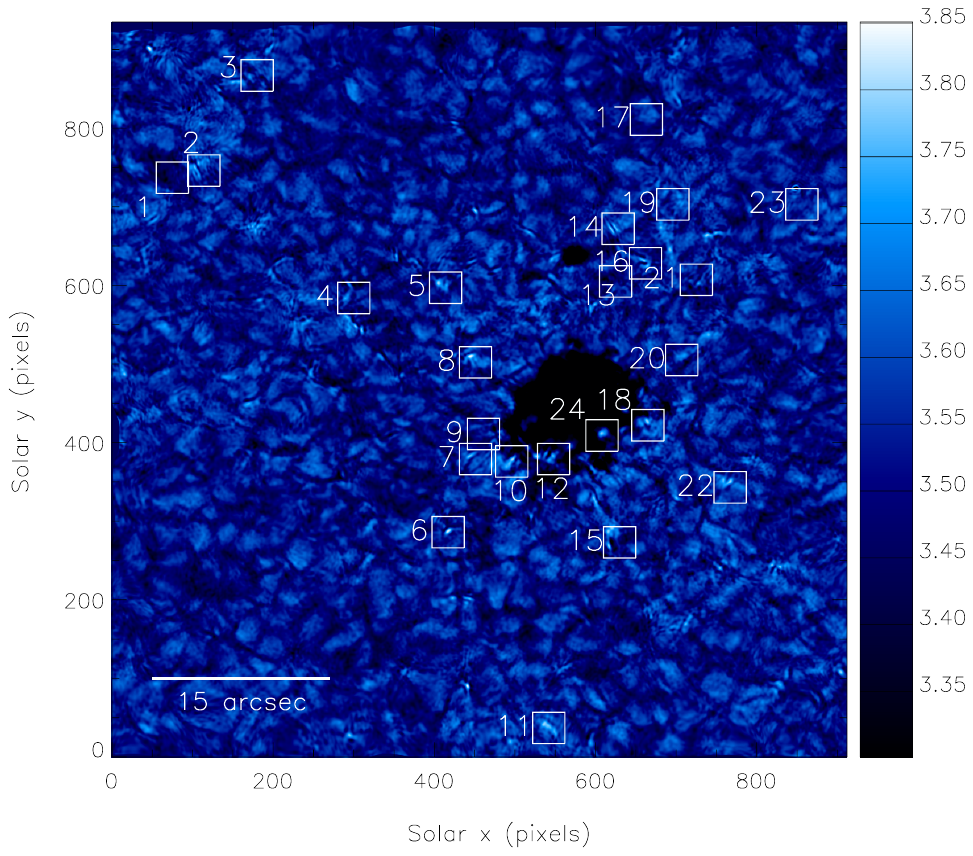} 
   \includegraphics[width=9cm]{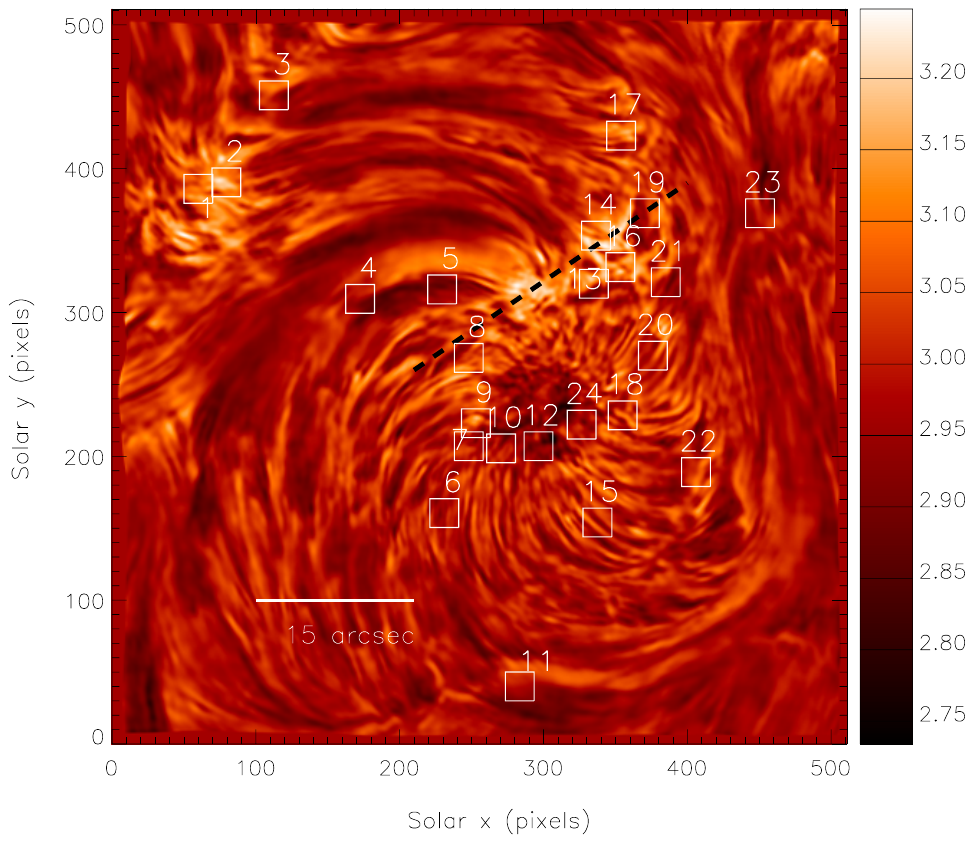} 
    \vspace{-0.5in}
      \caption{From left to right: Reconstructed ROSA images for the G-band and H$\alpha$ of active region NOAA~AR11249. The overlying boxes show the corresponding MBPs.  The dashed black line in the H$\alpha$ images (just above the AR) indicates the region used for the time-distance plot shown in Fig. \ref{Fig4XT}.
              }
         \label{FigVibStab}
   \end{figure*}

 Observations of the decaying active region NOAA~AR11249 (heliocentric coordinates S16.3, E03.3) were obtained with the ROSA \citep{2010SoPh..261..363J} and HARDCam \citep{2012ApJ...757..160J} imaging instruments on 2011~July~11 using the Dunn Solar Telescope in Sacramento Peak, USA.  
 Observations were conducted between 14:36 and 15:34~UT, with ROSA sampling the photosphere through a 9.20{\AA} (full width half maximum) G-band filter, while HARDCam obtained chromospheric observations through a 0.25{\AA} (full width half maximum) H$\alpha$ filter. The ROSA and HARDCam observations employed $0{\,}.{\!\!}''069$ and $0{\,}.{\!\!}''138$ per pixel spatial samplings, respectively to obtain field of view sizes of $69{\,}.{\!\!}''3 \times 69{\,}.{\!\!}''1$ and $71{\,}.{\!\!}''0 \times 71{\,}.{\!\!}''0$. 
 The exposure time for the G-band was 5 ms and was run at a cadence of 30.3 frames per second (fps). 
 The narrowband H$\alpha$ filter was run at a 27.9 fps cadence with a 35 ms exposure time. 
 High-order adaptive optics \citep{rimmele2004plasma} were utilized to correct wavefront deformations in real time during these observations and the images were speckle-reconstructed using $32 \rightarrow 1$ and $35 \rightarrow 1$ restorations 
 \citep{weigelt1983image, woger2008speckle} 
 for the G-band and H$\alpha$, respectively.  The reconstructed ROSA images for this study are shown in Fig. \ref{FigVibStab}
 with the bright points used in the study labeled. 
 We note that over the course of the 58-minute observation window, MBPs were detected within the boxes labeled 1--24. However, due to the image being a snapshot from a single point in time, it is possible that not all boxes simultaneously had MBPs present inside them. Furthermore, the MBPs have less contrast when observed in H$\alpha$, making them more difficult to visually identify in Fig. \ref{FigVibStab} \citep[similar to the effects noted by][]{samanta2016effects}.

\begin{figure}
\centering
\vspace{-0.3in}
\hspace*{-0.3cm}
\includegraphics[width=0.52\textwidth]{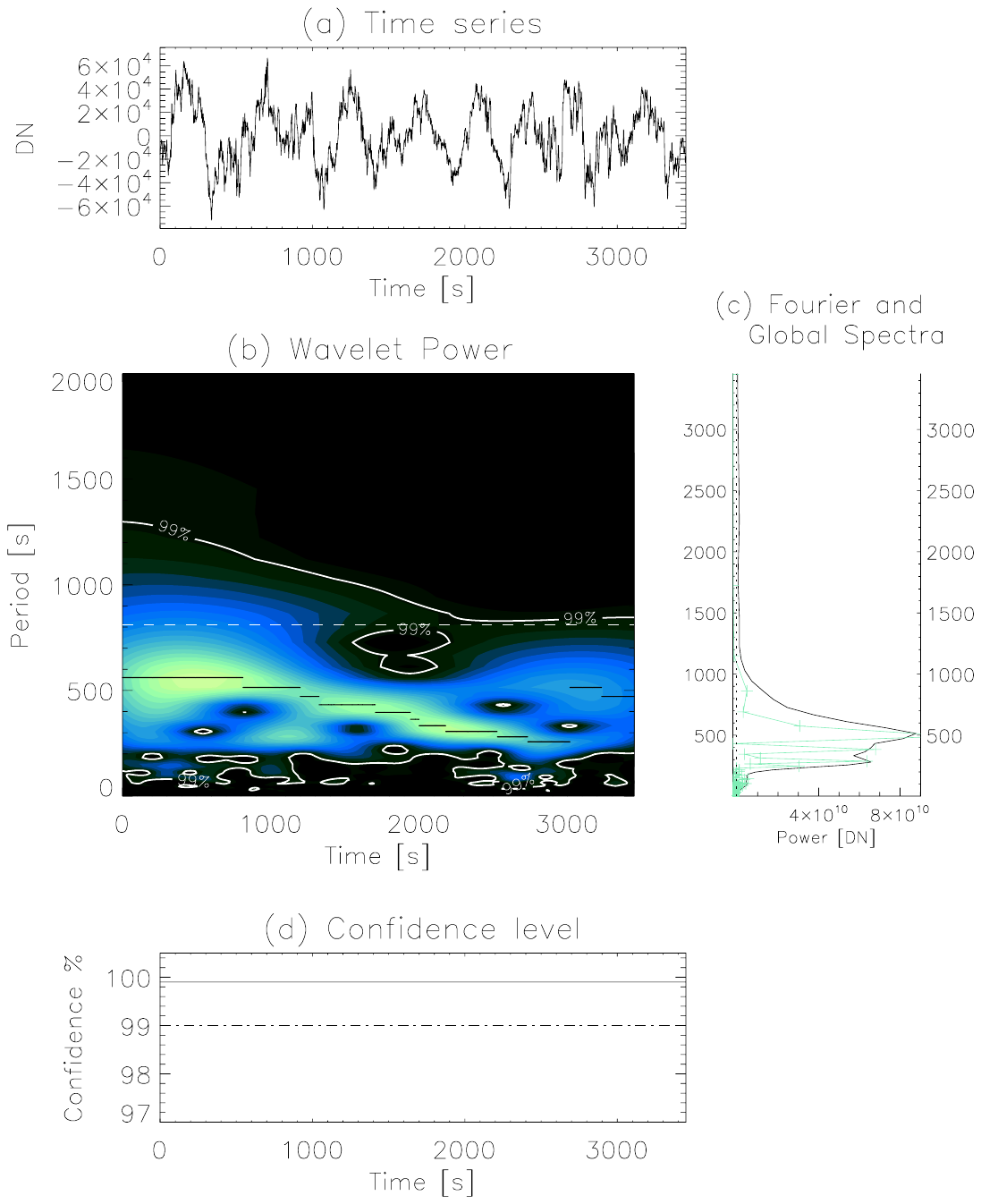}
\caption{Temporal analysis of a sample G-band light curve. (a) An MBP's de-trended G-band light curve at 541, 411 (MBP \#10; see Table \ref{FigVibStab}). 
(b) Wavelet power transform along with
locations at which detected power is at or above the 99\% confidence level contained within the contours.
(c) Summation of the wavelet power transform over time (full line) and
the fast Fourier power spectrum (crosses) over time, plotted as a function of period. Both methods show
strong detection near 540 seconds. (d)
Global wavelet (solid line) and Fourier (dashed-dotted
line) 99\% significance levels.  
}
         \label{Figwavelet}
   \end{figure}

 We employed a MBP tracking algorithm 
\citep{crockett2010tracking, Keys2020tracking} to select MBPs from the G-band images.
From the several hundred MBPs identified, we selected two dozen of the brightest MBPs, that is, with peak intensities over 2000 counts 
above the quiescent level for further temporal analysis.  
G-band MBPs were then matched to the co-aligned H$\alpha$ images. 
We derived frequencies and periods for each co-spatial 
bandpass using the wavelet analysis techniques described by
\cite{2012ApJ...757..160J}, first introduced by \cite{torrence1998practical}. Then, light curves were extracted \citep{Christian2019} for the G-band and H$\alpha$ datasets for each MBP.
Spatial regions of 40×40 and 20×20 pixels$^{2}$ were chosen for the G-band and H$\alpha$, respectively. We note that the H$\alpha$ images have a spatial sampling that is twice that of the G-band images, and hence 20×20 pixels$^{2}$ in H$\alpha$ occupies the same surface area as 40×40 pixels$^{2}$ in the G-band, which helps keep the data processing consistent between the bandpasses. Most examples of MBPs show photospheric and chromospheric signatures that are approximately co-spatial. However, some small spatial offsets were identified in several MBPs, such as the MBP \#11 near the edge of the detector, which may be a consequence of more heavily inclined magnetic field geometries in this location \citep[e.g.,][]{keys2013tracking}.
Each light curve was then de-trended by a first-order polynomial to remove long-term variations in intensity and normalized to their subsequent mean. Wavelet and Fourier analysis was then performed on each MBP light curve. An example of the timing analysis is shown in Fig. ~\ref{Figwavelet}.


%
%

\section{Results}
Using our wavelet analysis techniques, we found strong periodic signals in the G-band and H$\alpha$ time series for individual MBPs. The G-band periods range from $\approx$280 to 670 s, and the H$\alpha$ periods range from 230 to 730 s, presented in Table \ref{table1}.  
Table \ref{table1} gives the MBP number, the G-band position in (x, y) coordinates, the G-band and H$\alpha$ frequencies and periods, the ratio of the H$\alpha$ to G-band frequency, and the Pearson correlation coefficient.  
Our periods correspond to frequencies between 1.5 and 3.6 mHz in the photosphere (G-band) and frequencies of 1.4 to 4.3 mHz in the chromosphere (H$\alpha$).
These G-band periods closely correspond to the well-known
p-mode oscillation of 3 to 5 minutes \citep{lites1995possible}.
In the next section, we compare these results to mode coupling theory and examine whether these waves can contain enough energy to heat the chromosphere.

\begin{table*}
\caption{\centering MBP frequencies and periods \label{table1}
}

\centering \begin{tabular}{lcccccccr}

\hline
 
  &   &  \multicolumn{2}{c}{G-band} &  \multicolumn{2}{c}{H$\alpha$} &  &  \\ 
 MBP & Position &  Freq & Period & Freq & Period & Freq. Ratio & R$_P$ & \\
No.  & & (mHz) & (s) & (mHz) & (s) & & (Pearson) &  \\
    \hline \hline
1	&	120,702	&	2.12	&	471.5	&	3.00	&	333.0	&	1.42	&  0.73 & 	\\ 
2	&	159,781	&	1.78	&	560.7	&	1.64	&	611.1	&	0.92	& 0.26 &  \\ 
3	&	225,902	&	1.64	&	611.4	&	2.31	&	432.1	&	1.41	&$-$0.28	&   \\ 
4	&	345,619	&	1.50	&	666.7	&	1.95	&	513.9	&	1.30	& 0.28	&   \\ 
5	&	459,632	&	1.78	&	560.7	&	1.38	&	726.8	&	0.77	& $-$0.61	&  \\ 
6	&	462,321	&	1.64	&	611.4	&	2.52	&	396.3	&	1.54	& 0.26	&  \\ 
7	&	496,414	&	2.75	&	363.5	&	2.75	&	363.4	&	1.00	& $-$0.86	&   \\ 
8	&	496,537	&	1.78	&	560.7	&	1.38	&	726.8	&	0.77	&	$-$0.12  &   \\ 
9	&	506,446	&	1.50	&	666.7	&	3.00	&	333.0	&	2.00	&	0.28  &   \\ 
10	&	541,411	&	1.95	&	514.1	&	3.89	&	256.9	&	2.00	& $-$0.50	&  \\  
11	&	587,104	&	2.31	&	432.3	&	1.64	&	611.1	&	0.71	& $-$0.05	&   \\ 
12	&	593,414	&	3.57	&	280.3	&	4.24	&	235.6	&	1.19	& 0.37	&     \\ 
13	&	670,640	&	2.12	&	471.5	&	3.89	&	256.9	&	1.83	& 0.03	&  \\ 
14	&	673,707	&	1.78	&	560.7	&	1.78	&	560.4	&	1.00	& $-$0.13	& \\ 
15	&	675,308	&	1.95	&	514.1	&	1.50	&	666.4	&	0.77	& $-$0.28	& \\ 
16	&	707,663	&	1.64	&	611.4	&	3.89	&	256.9	&	2.38	& 0.45	&  \\ 
17	&	708,846	&	1.78	&	560.7	&	2.75	&	363.4	&	1.54	& 0.52	& \\ 
18	&	710,457	&	1.64	&	611.4	&	1.78	&	560.4	&	1.09	& 0.39	&  \\ 
19	&	741,738	&	1.95	&	514.1	&	2.31	&	432.1	&	1.19	& 0.16	&  \\ 
20	&	752,540	&	1.64	&	611.4	&	4.24	&	235.6	&	2.59	& $-$0.005	&  \\ 
21	&	770,642	&	2.12	&	471.5	&	4.24	&	235.6	&	2.00	& $-$0.24	&   \\ 
22	&	812,378	&	2.52	&	396.4	&	2.52	&	396.3	&	1.00	& 0.002	&  \\ 
23	&	901,738	&	1.95	&	514.1	&	1.95	&	513.9	&	1.00	& $-$0.26	& \\ 
24 & 653,444 & 1.78 & 560.7 & 3.27 & 305.6 & 1.83 & $-$0.42 \\  
 
 
 \hline
\end{tabular}
\\
\end{table*}

 \begin{figure*}[t]
   \sidecaption
   \includegraphics[width=0.51\textwidth]{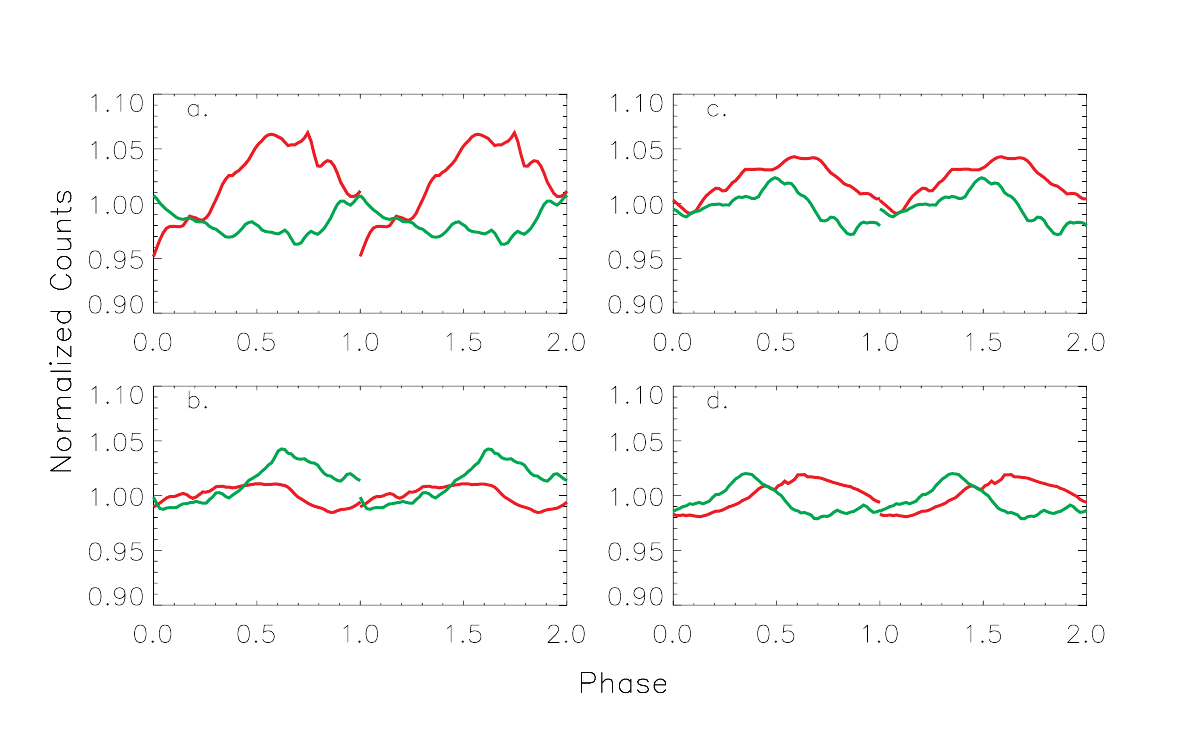} 
   \caption{Sample folded light curves comparing the G-band and H$\alpha$ as a phase function, each light curve is folded on its respective periods. The MBPs for the G-band positions are shown as: a. 459, 632 (MBP \#5); b. 670, 640 (MBP \#13); c. 708, 846 (MBP \#17); and d. 770, 642 (MBP \#21).
      Signals for panel a. show a strong anti-correlation.
      Signals from panel c. show a strong correlation, while panels b. and c. show weaker positive correlations (see the main text).\\\\\\\\
              }
         \label{Fig3fold}
   \end{figure*}
   
We compare the phases of the G-band and H$\alpha$ folded light curves and sample folded light curves in Fig. 3. The G-band and H$\alpha$ light curves were folded on their respective periods, and the cycle-to-cycle intensity variations were averaged to compute a total average phased light curve. We then computed the correlation of these light curves using the Pearson correlation coefficient (R$_P$). These values are given in Table \ref{table1}. Generally, the G-band and H$\alpha$ folded light curves show weak correlations, with an average Pearson value of $\approx$0.04. However, several bright points, 1, 12, 16, 17, and 18
show moderate to strong positive correlations, and bright points 5, 7, and 10 show strong negative correlations.  Additionally, 16 of the 24 MBPs show negative correlations. We compare these results to mode coupling theory in the Discussion.

\section{Discussion}
 
\subsection{Observed period and frequencies}
Many previous studies have shown that the highest concentrations of power are found in highly magnetized regions,
such as MPBs and intergranular lanes \citep{2023LRSP...20....1J}.
\cite{jess2007high} identified 20 and 370 s periods 
in both the G-band and H$\alpha$ blue wing, with a significant concentration of high-frequency activity ($>$ 20 mHz) in the sunspot penumbra.  
We find the period for our G-band MBPs
in the range 280 to 667 s (1.5 to 3.6 mHz).  
\cite{stangalini2015non}
discovered kink waves amplified in the 2-−8 mHz band, with a maximum of approximately 4.2 mHz in a study of 35 MBPs
that used the photospheric Fe I doublet at 630 nm and the Ca II H core line at 396.9 nm to cover the chromosphere. This range of frequencies covers what we observed in the G-band and H$\alpha$. 
Although not in the G-band, \cite{lites1993dynamics} found periods of 3 to 7 mHz in the photosphere. 
The H$\alpha$ periods as we approach the chromosphere tend to be much lower, with \citet{mcateer2003observational} finding significant power down to 230 s (4.3 mHz). We find a range of periods for the MBPs in H$\alpha$ (chromosphere) of 235 to 726 s
(1.4 to 4.3 mHz). 
\cite{morton2012observations} identified longitudinal (sausage) modes in H$\alpha$ with periods of 197 s.
Our typical H$\alpha$ period is 435 s, or 2.3 mHz, with the lowest periods being about 235 s (4.3 mHz), which is consistent with the McAteer value.

\subsection{Comparison to mode coupling theory}
Theory has shown that transverse (kink) waves created in the photosphere can be converted into longitudinal
waves at twice the frequency. 
\citet{ulmschneider1991propagation} found that monochromatic transverse waves of period P$_k$ transferred energy mainly 
to longitudinal waves of half the period, P$_l$ = P$_k$/2, but also transferred some energy, although much less, to waves with 
the period P$_l$ = P$_k$.
Several observational studies have found supporting evidence for this mode coupling. 
\citet{mcateer2003observational} found longitudinal waves 
(with frequencies between 2.6 and 3.8 mHz) 
in the chromosphere and transverse (kink) waves in the photosphere (with frequencies between 1.3 and 1.9 mHz). 
This was clear evidence of mode coupling with the longitudinal wave at twice the 
frequency of the transverse wave.  Additional studies by \cite{morton2012observations} found coupling of kink and sausage waves in the chromosphere. These waves were out of phase, possibly showing nonlinear resonance conditions. 
In a study of Type I spicules, \citet{jess2012chromo} found a form of “reverse” mode coupling where transverse oscillations 
found in the chromosphere with H$\alpha$ (periods of 65 to 220 s) were linked to longitudinal waves 
in the photosphere (periods of 130 to 440 s). Here, we find G-band frequencies ranging from 
1.5 to 3.6 mHz and H$\alpha$ frequencies ranging from 1.4 to 4.3 mHz. 
Our range of H$\alpha$ frequencies covers the range found by \cite{mcateer2003observational}, and 
we find that 9 of 24 MBPs (38\%) have ratios of H$\alpha$ to G-band frequencies greater than 1.5, consistent 
with predictions of mode coupling. An additional 12 of the 24 MBPs had ratios closer to 1, consistent with the lesser mode \citep{ulmschneider1991propagation}. 
We determined the wave types based on the ratios of H$\alpha$ to G-band frequencies observed in simple mode coupling theory.  We do not observe strong wave patterns \citep[such as that seen in][in time-distance plots]{jess2012chromo}
as is evident in our time-distance plot in Fig. \ref{Fig4XT}, nor do we have velocity information. 
Thus, we have little information for determining the wave nature. Future studies will have to wait for higher spatial resolution observations, such as those available with the {\textit{Daniel K. Inouye} Solar Telescope (DKIST)}. 

\subsection{Phases}
Simple mode coupling theory \citep{kalkofen1997oscillations} suggests the transverse (kink) waves formed in the photosphere are in phase with the longitudinal (sausage) waves in the chromosphere. \cite{lites1995possible} found strong phase coherence for frequencies near 5 mHz between these two waves, with all phase coherence disappearing at the highest frequencies.   We find more than half of our MBPs (13/24) show in-phase oscillations between the G-band (photosphere) and H$\alpha$ (chromosphere) with transverse and longitudinal wave types, respectively. However, phases of the H$\alpha$ and G-band light curves show strong positive and negative correlations only 21\% and 12\% of the time, respectively. Most of these in-phase MBPs have ratios of H$\alpha$ to G-band frequencies greater than 1, supporting simple mode coupling theory. The other MBPs are out of phase and generally have frequency ratios of less than 1.
Only four MBPs 
show strong positive correlations when comparing the phases of the G-band and H$\alpha$ folded light curves. This lack of strong correlations may be a function of poor matching from the G-band to H$\alpha$ bands, as features in the H$\alpha$ are more difficult to trace. 

\subsection{Energy budget}
One of the important aspects of detecting mode coupling between the photosphere and chromosphere is determining 
whether the longitudinal waves produce sufficient energy to heat the outer solar atmosphere.  
\cite{jess2009alfven} provided the first observational evidence of the torsional (Alfvén) waves, detected as 
full width half maximum oscillations in a small MBP through the lower solar atmosphere (with periods on the 
order of 126 to 700 s). They estimated an energy flux of $\approx$15,000 W m$^{-2}$ carried by these waves.
Similar to our study, ROSA 
observations of MHD waves by \cite{morton2012observations} found sufficient energy to heat the corona with an estimated 
energy flux for the incompressible fast kink mode of 4,300 $\pm$ 2,700 W m$^{-2}$ and an estimated wave energy flux 
for the compressible fast MHD sausage mode of 11,700 $\pm$ 3,800 W m$^{-2}$. 
 
\cite{jess2009alfven} estimated the energy flux for an Alfv\'en wave
as:
\begin{equation} 
\\\\ E = \rho v_{p}^2 v_A
\end{equation}

\begin{figure}
   \vspace{-0.9in}
   \hspace*{-0.7cm}
   \centering
   \includegraphics[width=10.5cm]{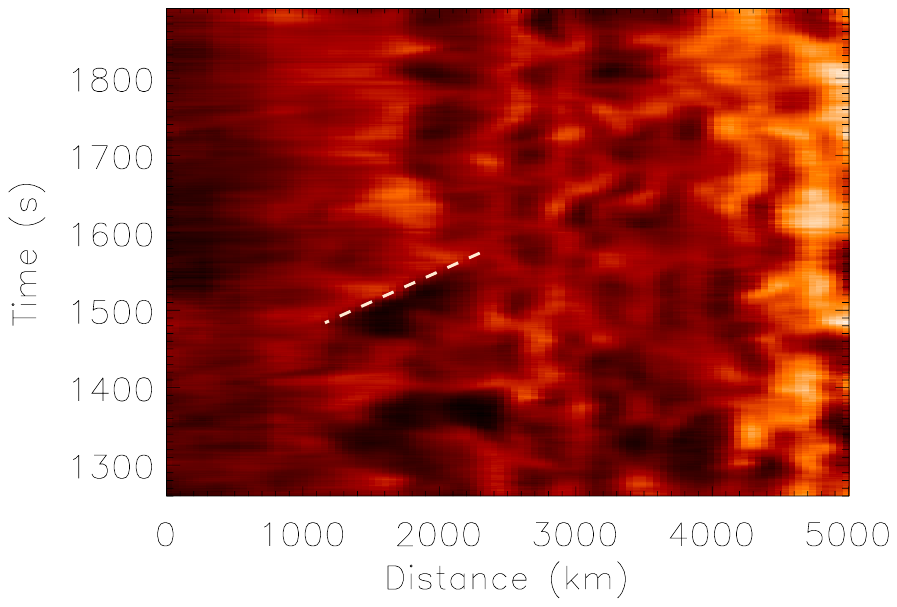}
   \vspace{-2.5in}
      \caption{Time-distance plot for H$\alpha$
       with the motion of a bright filament indicated by the dashed white line and moving at approximately 12.6 km/s. The extraction region from above the sunspot is indicated in the H$\alpha$ image in Fig. \ref{FigVibStab}b
       (see the main text).
              }
         \label{Fig4XT}
   \end{figure}

Where $\rho$ is the mass density, and v$_p$ and v$_A$ are the phase and Alfvén velocities, respectively. It is extremely difficult to estimate or infer all three of these plasma quantities simultaneously to a high degree of precision, and our energy flux estimate is an approximation. The density is normally taken from analytical models, such as the 1.3 $\times 10^{-8} \, kg{\,}m^{-3}$ used in \citet{jess2012chromo},
although modern inversion routines may be better able to estimate this quantity using full Stokes polarimetry (e.g., Hazel \citep{Hazel2008}; STIC \citep{STIC2019}; NICOLE \citep{Nicole2015}; and DeSIRe \citep{DESIRE2022}). 
Similarly, the Alfvén velocity requires an understanding of the torsional velocity amplitudes of the embedded wave, which is only possible through spectroscopic means \citep[e.g., the 22 km/s estimated by][]{jess2009alfven}. Typically, the phase velocity of the wave is a more straightforward property to estimate, since it can be linked to either the phase lag between signatures seen at different heights of the solar atmosphere \citep[e.g.,][]{jess2012cpropagating}
or the physical tracking of wave motion captured using time-distance seismology
\citep[e.g.,][]{morton2012observations,krishnaprasad2019}.
In our present work, Fig. \ref{Fig4XT} shows a time-distance diagram highlighting dynamical motion with velocities on the order of 12.6 km/s. In reality, the true velocity may be higher than 12.6 km/s due to inclination effects with respect to the observer's line of sight, although this still provides a lower estimate of the chromospheric dynamical velocities associated with MBPs. We note that models for flux tubes emerging through the photosphere show their horizontal velocity growing rapidly with height \citep{ulmschneider1991propagation}.
Thus, we expect there to be some horizontal component to the flux tube for heating as it reaches the 
chromosphere. Using this measured velocity (Fig. \ref{Fig4XT}), alongside the density and Alfvén velocity values of $\rho$ = 1.3 $\times 10^{-8}$ $kg{\,}m^{-3}$ and v$_A$ = 22 km/s, respectively, we calculated 
an energy flux of $\approx$45 $\times$ 10$^{3}$\,W{\,}m$^{-2}$. We note that this energy flux is sufficient to heat localized areas of the chromosphere and agrees with the energy estimates put forward by \citet{erdelyi2007theory}.

\cite{yadav2021slow} also found that longitudinal waves supply enough energy to heat the chromosphere in the solar plage. 
Thus, based on our work, the idea of mode coupling remains as a process for heating the solar chromosphere and corona remains viable. Future studies that include velocity information, such as those available with the new DKIST instruments, can further our understanding of the coronal heating issue.

\section{Conclusion}
We have used high temporal and spatial resolution observations of the 
solar photosphere and chromosphere to search for mode coupling in the solar atmosphere. Two dozen MBPs were studied and found to have frequencies between 1.5 and 3.6 mHz in the photosphere (G-band) and frequencies of 1.4 to 4.3 mHz in the chromosphere (H$\alpha$).
About 38\% of the MBPs  
show the expected frequency doubling for transverse waves in the photosphere converted into longitudinal waves in the chromosphere. More than half of our MBPs (13/24) show in-phase oscillations between the G-band (photosphere) and H$\alpha$ (chromosphere) with transverse and longitudinal wave types, respectively.
We find an energy flux of $\approx$45 $\times 10^{3}$ W m$^{-2}$, in agreement with 
previous studies which show that this mode coupling produces high enough energy fluxes to heat the solar chromosphere.
Future high temporal and spatial resolution studies that include velocity information, such as those available with the new 
DKIST instruments are needed to further our understanding of wave propagation and heating in the solar atmosphere.

\begin{acknowledgements}
We thank the CSUN Department of Physics and Astronomy for supporting this project. We acknowledge partial support of this project from NASA grants 19-HSODS-0004 and 21-SMDSS21-0047. We thank D. Gilliam and the Dunn Solar Telescope staff for their excellent support of the observations for this project. We wish to acknowledge scientific discussions with the Waves in the Lower Solar Atmosphere (WaLSA; \href{https://WaLSA.team}{www.WaLSA.team}) team, which has been supported by the Research Council of Norway (project no. 262622), The Royal Society \citep[award no. Hooke18b/SCTM;][]{2021RSPTA.37900169J}, and the International Space Science Institute (ISSI Team~502).
D.B.J. acknowledges support from the Leverhulme Trust via the Research Project Grant RPG-2019-371.
D.B.J. wishes to thank the UK Science and Technology Facilities Council (STFC) for the consolidated grants ST/T00021X/1 and ST/X000923/1.
D.B.J. and P.H.K. also acknowledge funding from the UK Space Agency via the National Space Technology Programme (grant SSc-009). We thank an anonymous referee for comments improving the manuscript. 
  
\end{acknowledgements}

\bibliographystyle{aa}
\bibliography{references}

\end{document}